\documentclass[a4paper,12pt]{article}
\usepackage{amsfonts}
\usepackage{amsthm}
\usepackage[top=2.75cm, bottom=3cm, left=2.75cm, right=2.75cm]{geometry}
\usepackage{multirow}
\title{The Large Vector Multiplet Action}
\author{Itai Ryb}

\newcommand{\beq}{\begin{equation}}
\newcommand{\eeq}[1]{\label{#1}\end{equation}}
\newcommand{\ber}{\begin{eqnarray}}
\newcommand{\eer}[1]{\label{#1}\end{eqnarray}}

\newcommand{\ft}[2]{{\textstyle\frac{#1}{#2}}}
\newcommand{\nll}{N\!=\!(1,1)}
\newcommand{\nZZ}{N\!=\!(2,2)}

\newcommand{\dq}[1]{\hat{q}^{#1}}
\newcommand{\jj}[3]{{J}_{#1 #3}^{#2}}
\newcommand{\bbD}[1]{\mathbb{D}_{#1}}
\newcommand{\bbDB}[1]{\bar{\mathbb{D}}_{#1}}
\newcommand{\bbG}[1]{\mathbb{G}_{#1}}
\newcommand{\bbGB}[1]{\bar{\mathbb{G}}_{#1}}
\newcommand{\bbX}[1]{\mathbb{X}_{#1}}
\newcommand{\bbXB}[1]{\bar{\mathbb{X}}_{#1}}
\newcommand{\bbXh}[1]{\hat{\mathbb{X}}_{#1}}
\newcommand{\bbXBh}[1]{\hat{\bar{\mathbb{X}}}_{#1}}

\newcommand{\hee}[1]{\Xi^1_{#1}}
\newcommand{\hcc}[1]{\Xi^2_{#1}}

\newcommand{\hcd}[1]{\nabla_{#1}}
\newcommand{\hPh}[1]{\varphi^{#1}}

\def\+{{+\!\!\!+}}

\newcommand{\aleq}{&\!\!\!=\!\!\!&}
\newcommand{\nn}{\nonumber}
\newcommand{\kah}{K\"ahler~}
\newcommand{\pa}[1]{\partial_{#1}}
\newcommand{\lam}{\Lambda}
\newcommand{\lamb}{\bar\Lambda}
\newcommand{\lamt}{\tilde\Lambda}
\newcommand{\lamtb}{\bar{\tilde{\Lambda}}}
\newcommand{\etc}{\textit{etc}}
\newcommand{\eg}{\textit{e.g.},~}
\newcommand{\cf}{\textit{c.f.},~}
\def\+{{+\!\!\!+}}

\begin{document}
\renewcommand{\theequation}{\thesection.\arabic{equation}}  
\setcounter{page}{0}
\thispagestyle{empty}

\begin{flushright} \small
YITP-SB-07-29\\ 
\end{flushright}
\smallskip
\begin{center} \LARGE
{\bf  The Large Vector Multiplet Action}
 \\[12mm] \normalsize
{\bf Itai Ryb} \\[8mm]
{\small\it
C.N.Yang Institute for Theoretical Physics, Stony Brook University, \\
Stony Brook, NY 11794-3840,USA\\}
\end{center}
\vspace{10mm}
\centerline{\bf\large Abstract} 
\bigskip
\noindent  
In this short note we discuss possible actions for the $d=2,~\nZZ$ large vector multiplet of \cite{Lindstrom:2007vc,Lindstrom:2007sq}. We explore two scenarios that allow us to write kinetic and superpotential terms for the scalar field-strengths, and write kinetic terms for the spinor invariants that can introduce topological terms for the connections.
\eject
\normalsize
\eject	
\section{Introduction}
Generalized \kah manifolds, which are torsionful manifolds equipped with two complex structures 
\cite{gualtieri}, arise as target space for $d=2,~\nZZ$ $\sigma$-models with both twisted chiral 
and chiral superfields \cite{Gates:1984nk}, or semichiral superfields \cite{Buscher:1987uw}. These 
manifolds and their world-sheet origins are subject to growing interest for both mathematicians 
and string theorists \cite{gualtieri,many}.

Recently, new $\nZZ$ multiplets were introduced \cite{Lindstrom:2007vc,Gates:2007ve} to gauge 
isometries in generalized \kah manifolds \cite{Merrell:2006py} and to show that the duality 
introduced in \cite{Grisaru:1997ep} is, in fact, T-duality \cite{Lindstrom:2007sq,Merrell:2007sr}; 
in particular it was shown that this T-duality relates (twisted) chiral multiplets $(\chi)\phi$ to 
semichiral multiplets $\bbX{L,R}$,
\begin{equation}
\label{eq::tduality}
K(\phi,\bar{\phi},\chi,\bar{\chi},x) 
\stackrel{\hbox{\tiny{T-duality}}}{\longleftrightarrow}\tilde{K} (\bbX{L} , \bbX{R} , \bbXB{L},
\bbXB{R},x)~,
\end{equation}
where $x$ are arbitrary spectator fields. In \cite{Lindstrom:2007vc} it was shown that the 
action for the semichiral vector multiplet (gauging the RHS of eq.~\ref{eq::tduality}) 
corresponds to that of one ordinary vector multiplet field-strength and three scalar 
multiplets. However, the large vector multiplet (gauging the LHS of eq.~\ref{eq::tduality}) 
has four extra spinor multiplets that can complicate the construction of 
kinetic terms by introducing higher derivative actions.

In this note we address this problem. We find two possible types of actions, 
one consisting of chiral field-strengths and the other consisting of twisted chiral field-strengths where higher derivative terms are explicitly eliminated using particular field redefinitions. We then discuss possible superpotentials that can accompany those kinetic terms, as well as actions for the spinor invariants, which are found to be field theories with first derivatives for both 
bosons and  fermions. Finally, we give the modified matter couplings due to the field redefinitions.

Throughout this paper we follow the notation of \cite{Lindstrom:2007sq}.

\section{Review: The large vector multiplet in $\nZZ$ and $\nll$ superspace}
We start our discussion with a \kah potential for (twisted) chiral superfields with a gauged 
isometry
\begin{equation}
\label{eq:kahlerpot}
K(i(\phi-\bar{\phi})+V^\phi~,~i(\chi-\bar{\chi})+V^\chi~,~ \phi+\bar{\phi} - 
\chi-\bar{\chi}+V^\prime )~;
\end{equation}
we use the notation of \cite{Lindstrom:2007sq} where the transformation properties for 
the three (real) superfields are
\begin{equation}
\delta V^\prime = -\lam -\lamb + \lamt + \lamtb ~~,~~~ \delta V^\phi = i(\lamb-\lam) ~~,~~~ 
\delta V^\chi = i (\lamtb-\lamt)~.
\end{equation}
These combine to give the complex potentials:
\begin{eqnarray}
V^L = \ft{1}{2}(-V^\prime + i V^\phi - i V^\chi) &\rightarrow& \delta V^L = \lam-\lamt~, \nn \\
V^R = \ft{1}{2}(-V^\prime + i V^\phi + i V^\chi) &\rightarrow& \delta V^R = \lam-\lamtb~,
\end{eqnarray}
along with their complex conjugates, which are potentials for the semichiral gauge-invariant field-strengths
\begin{equation}
\bbG{+} = \bbDB{+} V^L ~~,~~~ \bbG{-} = \bbDB{-} V^R ~~,~~~ \bbGB{+} = \bbD{+} \bar{V}^L ~~,~~~ 
\bbGB{-} = \bbD{-}\bar{V}^R ~,
\end{equation}
and eight chiral and twisted chiral field-strengths:
\begin{eqnarray}
W       = i \bbDB{+}\bbDB{-}V^\chi &,&
B       = i \bbDB{+}\bbDB{-}(-V^\prime + i V^\phi)~, \nn \\
\bar{W}     = i \bbD{+}\bbD{-}  V^\chi &,&
\bar{B}     = i \bbD{+}\bbD{-}  (-V^\prime - i V^\phi) ~, \nn \\
\tilde{W}   = i \bbDB{+}\bbD{-} V^\phi &,&
\tilde{B}   = i \bbDB{+}\bbD{-} (-V^\prime - i V^\chi) ~, \nn \\
\bar{\tilde{W}} = i \bbD{+}\bbDB{-} V^\phi &,&
\bar{\tilde{B}} = i \bbD{+}\bbDB{-} (-V^\prime + i V^\chi) ~.
\end{eqnarray}
These can be expressed as $\nZZ$ supercovariant derivatives on the invariant spinors:
\begin{eqnarray}
W      = + (\bbDB{+}\bbG{-}  + \bbDB{-}\bbG{+}) &,&
B      = i (\bbDB{+}\bbG{-}  - \bbDB{-}\bbG{+}) ~, \nn \\
\bar{W}    = - (\bbD{+} \bbGB{-} + \bbD{-}\bbGB{+}) &,&
\bar{B}    = i (\bbD{+} \bbGB{-} - \bbD{-}\bbGB{+}) ~, \nn \\
\tilde{W}  = - (\bbDB{+} \bbGB{-} + \bbD{-}\bbG{+}) &,&
\tilde{B}  = i (\bbDB{+} \bbGB{-} - \bbD{-}\bbG{+}) ~, \nn \\
\bar{\tilde{W}}= + (\bbD{+} \bbG{-} + \bbDB{-}\bbGB{+}) &,&
\bar{\tilde{B}}= i (\bbD{+} \bbG{-} - \bbDB{-}\bbGB{+}) ~.
\end{eqnarray}
The descent to $\nll$ uses the decompositions of the $\nZZ$ derivatives and fields:
\begin{equation}
\bbD{\pm} = \ft{1}{2}(D_\pm - i Q_\pm) ~~\hbox{and}~~ \left.\bbG{\pm}\right| = \hee{\pm} + i \hcc{\pm}
\end{equation}
where $|$ indicates projection to $\nll$. We use a real basis of $\nll$ gauge-invariant fields \cite{Lindstrom:2007sq}:
\begin{eqnarray}
\dq{\phi} = -i\ft{1}{2} (Q_{[+}\hee{-]}-D_{[+}\hcc{-]}) &,& iD_\pm \Xi_\mp^{1,2}  ~,\nn \\
 \dq{\chi} = -i\ft{1}{2}(Q_{(-}\hee{+)}+D_{(+}\hcc{-)}) &,& \dq{\prime} = -i\ft{1}{2} Q_{[+}\hcc{-]} ~,
\end{eqnarray}
as well as the field-strength for the $\nll$ connections $A_\pm = \ft{1}{4}Q_\pm 
(V^\phi \pm V^\chi)$: 
\begin{equation}
f=-i(Q_+ \hcc{-}+Q_- \hcc{+}) = i (D_+ A_- + D_- A_+)~.
\end{equation}
The linear relations between the $\nZZ$ and $\nll$ invariants could be summarized in the matrix equation:
\begin{equation}
\label{eq::linear_relation}
L= \left. \left( \begin{array}{c}
W \\
B \\
\bar{W} \\
\bar{B} \\
\tilde{W} \\
\tilde{B} \\
\bar{\tilde{W}} \\
\bar{\tilde{B}}
\end{array} \right)\right| = \frac{1}{2} \left( \begin{array}{rrrrrrrr}
-i&-i& 0& 0& 0&-1& 0&-i\\
 1&-1& i&-i&-i& 0& 1& 0\\
 i& i& 0& 0& 0&-1& 0& i\\
 1&-1&-i& i& i& 0& 1& 0\\
 i& i& 0& 0& 1& 0& 0&-i\\
 1&-1&-i&-i& 0&-i&-1& 0\\
-i&-i& 0& 0& 1& 0& 0& i\\
 1&-1& i& i& 0& i&-1& 0\\
\end{array}\right) \left( \begin{array}{c}
iD_+ \hee{-} \\
iD_- \hee{+} \\
2iD_+ \hcc{-} \\
2iD_- \hcc{+} \\
2\dq{\phi}\\
2\dq{\chi}\\
2 \dq{\prime}\\
f
\end{array}
\right) = UL^{\prime}
\end{equation}
\section{Possible candidates for large multiplet action}
\subsection{Na\"ive kinetic terms for $\dq{\prime},\dq{\phi,\chi}$ and $f$}\label{ssec::kinetic_term}
A generic, Lagrange density in the (twisted) chiral invariants has eight free 
parameters\footnote{Terms mixing chiral and twisted chiral fields lead to total derivatives and could be interesting for global considerations.}. These correspond to two $2\times 2$ Hermitian matrices $s_{c,t} = s_{c,t}^\dagger$:
\begin{equation}
\label{eq::kinetic}
\mathcal{L}_{kin} =
(W,B) s_c \left( \begin{array}{c} \bar{W} \\ \bar{B} \end{array} \right) +
(\tilde{W},\tilde{B}) s_t
\left( \begin{array}{c} \bar{\tilde{W}} \\ \bar{\tilde{B}} \end{array} \right) = L^i L^j S_{ij} 
\end{equation}
Reduction of such a density to $\nll$ is straightforward using the matrices $S$ (eq. \ref{eq::kinetic}), $U$ (eq. \ref{eq::linear_relation}), and the complex structures $J_\pm = \hbox{diag}(i,-i,\pm i , \mp i)$; after integration by parts we find:
\begin{eqnarray}
\int D_+ D_- Q_+ Q_- \mathcal{L}_{kin} \aleq 2 \int D_+ D_- \big( U^T (J_+ S J_- - J_+ J_- S )U\big)_{ij} 
D_+ L^{\prime i} D_- L^{\prime j} \nn \\
\aleq \int D_+ D_- S^{\prime}_{ij}D_+ L^{\prime i} D_- L^{\prime j}~.
\end{eqnarray}

Terms of the form $D_{[\pm]}D_\pm\Xi_\mp^A$ lead to second derivatives 
on spinors and are, therefore, to be avoided. An Ansatz that leads to this desired 
result involves a particular choice of linear field redefinitions:
\begin{eqnarray}
&& L^{\prime \prime} = U^{\prime} L ~~,~~~ U^{\prime} = \left( \begin{array}{cccc}
 1 & 0 \\
\alpha & 1
\end{array} 
\right) U~~,~~~ L^{\prime\prime}_{1,2,3,4}=0 \nn \\
&& A_\pm \rightarrow A_\pm + \alpha^{(\pm)}_{B} \Xi^B_\pm ~~,~~~ \dq{i} \rightarrow \dq{i} + \beta^{i(\pm)}_{B} 
D_\pm \Xi^B_\mp ~~,~~~ B=1,2
\end{eqnarray}
and parameters $s_{c,t}$\footnote{It is impossible to cancel all higher derivative terms solely by adjusting the parameters $s_{c,t}$.} Where all blocks of are $4 \times 4$. 

The matrix $U^\prime$ is invertible and we therefore require, to propagate all field-strengths, that $J_+ S J_- - J_+ J_- S$ has four vanishing eigenvalues. Diagonalizing $S$ we find four pair of eigenvalues, two pairs due to $s_c$ and two pairs due to $s_t$ that brings us to the following classes of consistent kinetic terms:
\begin{itemize}
 \item Chiral: $s_t=0 ~,~ s_c$ is arbitrary hermitian.
 \item Twisted chiral: $s_c=0 ~,~ s_t$ is arbitrary hermitian.
\item Mixed: $\det s_c = \det s_t = 0 $
\end{itemize}

One can verify by explicit substitution of the blocks of $S^{\prime}$ \begin{equation}
\label{eq::S_PRIME_BLOCK}
S^{\prime} = \left( \begin{array}{cc}
 A & B \\
B^T & C
\end{array}
 \right)
\end{equation}
that these choices indeed satisfy the condition for consistent field redefinition
\begin{equation}
\label{eq::field_redefinition_condition}
\left( \begin{array}{cccc}
 1 & \alpha^T \\
0 & 1
\end{array}
\right) \left( \begin{array}{cc}
 A & B \\
B^T & C
\end{array}
 \right) \left( \begin{array}{cccc}
 1 & 0 \\
\alpha & 1
\end{array}
\right) = \left( \begin{array}{cc}
 0 & 0 \\
0 & C
\end{array}
\right) \rightarrow A - BC^{-1}B^T = 0~.
\end{equation}

Analysis of the nonabelian extension to the large vector multiplet introduces further restrictions on the large vector multiplet action \cite{Lindstrom:2008hx}, allowing only the mixed solution and restricting to (twisted) chiral combinations of the form $W\pm i B$ \etc. We now present in detail the classes unique to the abelian case: the chiral and the twisted chiral solutions. The field redefinitions are listed in table 
\ref{tab:scenarios}.

\begin{table}[h!]
\caption{Field redefinitions for kinetic terms 
\label{tab:scenarios}}
$$ \begin{array}{|c||c|c|}
\hline &&\\[-3mm]
&\hbox{Chiral} &\hbox{Twisted chiral}\\[1mm]
~& s_t = 0 & s_c = 0 \\[1mm] \hline \hline &&\\[-3mm] 
A_\pm \rightarrow &A^{(c)}_\pm = A_\pm + \hee{\pm}& A^{(t)}_\pm = A_\pm - \hee{\pm} \\[1mm] \hline &&\\[-3mm]
f \rightarrow &f_c = f + iD_{(+}\hee{-)}&f_t = f - iD_{(+}\hee{-)} \\[1mm] \hline &&\\[-3mm]
\dq{\prime} \rightarrow &\dq{\prime}_c = \dq{\prime} + \ft{1}{2}iD_{[+} \hee{-]}&\dq{\prime}_t = \dq{\prime} - \ft{1}{2}iD_{[+} \hee{-]} \\[1mm] \hline &&\\[-3mm]
\dq{\chi} \rightarrow &\dq{\chi}_c = \dq{\chi} &\dq{\chi}_t =  \dq{\chi}+iD_{(+}\hcc{-)} \\[1mm] \hline &&\\[-3mm]
\dq{\phi} \rightarrow & \dq{\phi}_c =\dq{\phi}-iD_{[+}\hcc{-]} & \dq{\phi}_t =\dq{\phi} \\[1mm] \hline
\end{array} $$
\end{table}

An explicit form for these kinetic terms after field redefinition in $\nll$ superspace is obtained by reduction:
\begin{itemize}
\item Chiral:\\
Writing the entries of $s_c$ explicitly
\begin{equation}
 s_c = s = \left( \begin{array}{cc}
 a & b+ ic \\
b-ic & d
\end{array}
\right)
\end{equation}
we push $Q_\pm$ through and find:
\begin{eqnarray}
&& \int Q_+ Q_- \left[ (W,B) s_c\!\left( \begin{array}{c} 
\bar{W} \\
\bar{B}
\end{array}
\right)\right] = \\
&&\ft{1}{2} \int D_+ (2\dq{\phi}_c,2\dq{\chi}_c, 2 \dq{\prime}_c,f_c) \left( \begin{array}{rrrr}
d & c& 0& b \\
c & a&-b& 0 \\
0 &-b& d& c \\
b & 0& c& a
\end{array}
\right) D_-\!\!\left( \begin{array}{c}
2\dq{\phi}_c\\
2\dq{\chi}_c\\
2\dq{\prime}_c\\
f_c
\end{array} \right)~.
\end{eqnarray}
where $\int$ is the $\nll$ measure $\int d^2 z D_+ D_-$. 
\item Twisted chiral:\\
In a similar fashion, we set the entries $s_t=s$ and reduce the twisted chiral action to $\nll$:
\begin{eqnarray}
&& \int Q_+ Q_- \left[ (\tilde{W},\tilde{B}) s_t\!\left( \begin{array}{c}
\bar{\tilde{W}} \\
\bar{\tilde{B}}
\end{array}
\right)\right] = \\
&&\ft{1}{2} \int D_+ (2\dq{\phi}_t,2\dq{\chi}_t,2\dq{\prime}_t,f_t) \left( \begin{array}{rrrr}
-a& c& b& 0\\
 c&-d& 0&-b\\
 b& 0&-d& c\\
 0&-b& c&-a
\end{array}
\right) D_-\!\!\left( \begin{array}{c}
2\dq{\phi}_t\\
2\dq{\chi}_t\\
2\dq{\prime}_t\\
f_t
\end{array} \right) ~. \nn
\end{eqnarray}
\end{itemize}

\subsection{Mass-like terms}
We now investigate terms of the form
\begin{equation}
\label{eq::mass_like}
\int i \bbDB{+} \bbDB{-} P_c (W,B)
+ \int i \bbDB{+} \bbD{-} P_t(\tilde{W},\tilde{B})
+ \int i \bbD{+} \bbD{-} \bar{P}_c (\bar{W},\bar{B})
+ \int i \bbD{+} \bbDB{-} \bar{P}_t(\bar{\tilde{W}},\bar{\tilde{B}})
\end{equation}
that arise naturally from a na\"ive mass term for the invariant semichiral spinors $\bbG{\pm}$, 
\eg 
\begin{equation}
\int \bbD{+}\bbD{-} \bbDB{+}\bbDB{-} \left( \bbG{+}\bbG{-} \right) = \int \bbD{+}\bbD{-} \left( 
\bbDB{-} \bbG{+} \bbDB{+} \bbG{-} \right) = \ft{1}{4} \int \bbD{+}\bbD{-} \left(W^2 + B^2 \right)~,
\end{equation}
and reduce to $\nll$ superspace in a straightforward manner due to the (twisted) chirality of the 
field-strengths \cite{Lindstrom:2007vc}.

We construct sensible candidates by taking into account the field redefinitions of section 
\ref{ssec::kinetic_term}, and requiring the absence of terms of the form
\begin{equation}
\label{eq:bad_term}
\int D_+ D_- \left( D_+ \Xi^A_- m_{AB} D_- \Xi^B_+ \right) ~,
\end{equation}
which give, when reduced to components, higher derivatives on spinors. As we now see, these terms 
can give:
\begin{itemize}
\item Superpotentials for  $\dq{\prime}_{c,t},\dq{\phi,\chi}_{c,t}$ and $f_{c,t}$.
\item Kinetic terms for $\Xi_\pm^A$ which are first order in derivatives.
\item Topological terms.
\end{itemize}

\subsubsection{Superpotentials for $\dq{\prime},\dq{\phi,\chi}$ and $f$}
After carrying out the field redefinitions of sec. \ref{ssec::kinetic_term} we have, in each 
scenario, four field-strengths that contain only the redefined $\dq{\prime}_{c,t},\dq{\phi,\chi}_{c,t}$ and $f_{c,t}$. We can, therefore, write in the chiral scenario any function for $P_c(W,B)$ (\cf eq.~\ref{eq::mass_like}) which 
reduces to $\nll$ superspace as:
\begin{equation}
2\int iD_+ D_- \hbox{Re} \left( P_c ( 2 \dq{\chi}_c + if_c , \dq{\prime}_c - i \dq{\phi}_c ) \right)~.
\end{equation}
In a similar fashion we write in the twisted chiral scenario a superpotential $P_t( \tilde{W} , 
\tilde{B} )$ (\cf eq.~\ref{eq::mass_like}) that reduces to $\nll$ superspace as:
\begin{equation}
2\int iD_+ D_- \hbox{Re} \left( P_t (2 \dq{\phi}_t - if_t ,  \dq{\prime}_t + i \dq{\chi}_t) \right)~.
\end{equation}
Particular examples of such superpotentials include mass and Fayet-Illiopoulos terms.

\subsubsection{Kinetic and topological terms for $\Xi_\pm^A$ \label{sssec::kinetic}}

After chiral field redefinition, we can write combinations of the twisted chiral field strengths:
\begin{eqnarray}
\tilde{W} + i \tilde{B} \aleq 2 i D_+ ( i\hee{-} + \hcc{-}) + \dq{\phi}_c + \dq{\chi}_c -i\dq{\prime}_c -\ft{i}{2} f_c ~,\nn \\
\tilde{W} - i \tilde{B} \aleq 2 i D_- (  i\hee{+} - \hcc{+} ) + \dq{\phi}_c - \dq{\chi}_c +i\dq{\prime}_c -\ft{i}{2} f_c ~,
\end{eqnarray}
that separate left and right spinors, and introduce a generic, 
\textit{``harmolomorphic''}\footnote{Since it obeys $\pa{\bar{\tilde{W}}} P_t = 
\pa{\bar{\tilde{B}}} P_t = \pa{\tilde{W}+i\tilde{B}}\pa{\tilde{W}-i\tilde{B}}P_t=0$.}, function for the twisted chirals:
\begin{equation}
P_t(\tilde{W},\tilde{B}) = P_{t+}(\tilde{W}+i\tilde{B})+P_{t-}(\tilde{W}-i\tilde{B})~,
\end{equation}
that contains no bad terms.

A quadratic function in $\tilde{W}\pm i \tilde{B}$ generates kinetic terms for the components 
of the spinors $\Xi^A_\pm$:
\begin{equation}
\left. \Xi^A_\pm \right|=\Xi^A_\pm ~~,~~~\left. D_{\pm}\Xi^A_\pm \right|=V_{\mbox{\tiny $ \stackrel\+ =$}}^A
~~,~~~\begin{array}{l}
  \left. D_{-}\Xi^A_+ \right|=b^A \\[1mm]
  \left. D_{+}\Xi^A_- \right|=b^{\prime A} \\
\end{array}
 ~~,~~~ \left. D_+ D_- 
\Xi^A_\pm \right|= \xi^A_\pm~.
\end{equation}
When reducing a term of the form, \eg $D_- \Xi^A_+ D_- \Xi^B_+$ appearing in 
$(\tilde{W}-i\tilde{B})^2$ we find\footnote{These terms can be interpreted as a twist of known ghost actions with $-3/2$ the ghost number \cite{Berkovits:1994vy}. I thank Martin Ro\v{c}ek and Cumrum Vafa for pointing this out.}
\begin{equation}
 D_+D_-\big( D_- \Xi^A_+ D_- \Xi^B_+ \big)=\pa{=} V_\+ ^{(A} b^{B)}+\xi_+^{(A} \pa{=}\Xi_+^{B)}~.
\end{equation}

Note that $(\tilde{W}\pm i \tilde{B})^2$ also introduce terms of the form $D_\pm \Xi^A_\mp f$, which, when reduced to components using
\begin{equation}
\left. f \right| = f ~~,~~~\left. D_\pm f \right| = \lambda_\pm ~~,~~~\left. D_+ D_- f \right| = F 
= \pa{\+}A_{=}-\pa{=}A_{\+} ~,
\end{equation}
gives terms such as $bF$ which are topological.

In the twisted chiral scenario we write the combinations
\begin{eqnarray}
W+iB \aleq 2iD_-(\hcc{+}-i\hee{+})+\dq{\phi}_t-\dq{\chi}_t+i\dq{\prime}_t-\ft{i}{2}f_t ~, \nn\\
W-iB \aleq 2iD_+(\hcc{-}-i\hee{-})-\dq{\phi}_t-\dq{\chi}_t-i\dq{\prime}_t-\ft{i}{2}f_t~,
\end{eqnarray}
for the chiral field-strengths which allows us to write kinetic terms for the spinor invariants such as
\begin{equation}
P_c(W,B) = P_{c+}(W+iB)+P_{c-}(W-iB)~.
\end{equation}

\section{Matter couplings revised}

In \cite{Lindstrom:2007vc,Lindstrom:2007sq} we discussed the reduction of the invariant \kah 
potential (\ref{eq:kahlerpot}) and wrote down the couplings of the large vector multiplet to (twisted) chiral matter. In particular, we wrote the gauge covariant derivative
\begin{equation}
\nabla_\pm \varphi^i = D_\pm \varphi^i - A_\pm k^i~~,~~~ \varphi^i = (\phi,\bar{\phi},\chi,\bar{\chi})~~,~~~ \mathcal{L}_k K = 0
\end{equation}
which couples $\varphi^i$ minimally to the connections. When including the action terms, the field redefinitions of section \ref{ssec::kinetic_term} modify the matter couplings. In the chiral scenario, we change
\begin{equation}
\nabla_\pm \hPh{i}- \hee{\pm} k^i = \nabla^{c}_\pm \hPh{i}
\end{equation}
and keep the form of the reduced Lagrange density\footnote{If the density is invariant only up to generalized \kah transformations we replace the Lie derivatives $(K_i  \jj{\pm}{i}{j} k^j ~,~ K_i \Pi^i{}_j k^j)$ with the moment maps $ (- \mu_{_\pm} ~,~ -\mu_{_\Pi}$) \cite{LRRvUZ}.}
\begin{eqnarray}
\label{eq::reduced_density}
\mathcal{L}_{m(c)} \aleq \left( \Xi_+^A + \hcd{+}^{c} \hPh{i} E_{iC} E^{CA} \right)
E_{AB} \left( \Xi_-^B + E^{BD} E_{Dj} \hcd{-}^{c} \hPh{j} \right) \nn\\[1mm]
&& + \hcd{+}^{c} \hPh{i} \left( E_{ij} - E_{iA} E^{AB} E_{Bj}\right) \hcd{-}^{c}
\hPh{j} \nn \\
&& + i K_i k^k \left( \dq{\phi}_c (\jj{+}{i}{k}+\jj{-}{i}{k}) + \dq{\chi}_c (\jj{+}{i}{k}-\jj{-}{i}{k}) + \dq{\prime}_c \Pi^i{}_k \right)~,
\end{eqnarray}
where we introduce the matrices
\begin{eqnarray}
E_{kl} \aleq K_{ij} \left( \jj{+}{i}{k}\jj{-}{j}{l} - \ft{1}{2}\Pi^i{}_k \delta^j{}_l - \ft{1}{2} 
\delta^i{}_k \Pi^j{}_l \right) \\[2mm]
E_{Al} \aleq K_{ij} \left( \begin{array}{c}
(\jj{+}{i}{k}+\jj{-}{i}{k}) \jj{-}{j}{l}-\Pi^i{}_k \delta^j{}_l \\[1mm]
 \Pi^i{}_k \jj{-}{j}{l} + (\jj{+}{i}{k}+\jj{-}{i}{k}) \delta^j{}_l 
\end{array}
\right) k^k \\[2mm]
E_{kB} \aleq K_{ij} \Big( \jj{+}{i}{k}(\jj{+}{j}{l}+\jj{-}{j}{l}) -\delta^i{}_k \Pi^j{}_l ~,~
\jj{+}{i}{k} \Pi^j{}_l + \delta^i{}_k (\jj{+}{j}{l}+\jj{-}{j}{l}) \Big) k^l \\[2mm]
E_{AB} \aleq K_{ij} \left( \begin{array}{cc}
(\jj{+}{i}{k} + \jj{-}{i}{k})(\jj{+}{j}{l}+\jj{-}{j}{l})  & (\jj{+}{i}{k}+\jj{-}{i}{k}) \Pi^j{}_l 
\\[1mm]
\Pi^i{}_k (\jj{+}{j}{l}+\jj{-}{j}{l})  & \Pi^i{}_k \Pi^j{}_l
\end{array} \right) k^k k^l
\end{eqnarray}
and $E^{AB}$ is the inverse of $E_{AB}$. 

In the twisted chiral scenario we change $(A_\pm^{(c)},\dq{\prime}_c,\dq{\phi,\chi}_c) \rightarrow (A_\pm^{(t)},\dq{\prime}_t,\dq{\phi,\chi}_t)$ and modify the matrices
\begin{eqnarray}
E_{Al} \aleq K_{ij} \left( \begin{array}{c}
-(\jj{+}{i}{k}-\jj{-}{i}{k}) \jj{-}{j}{l}+\Pi^i{}_k \delta^j{}_l \\[1mm]
 \Pi^i{}_k \jj{-}{j}{l} + (\jj{+}{i}{k}-\jj{-}{i}{k}) \delta^j{}_l
\end{array}
\right) k^k \\[2mm]
E_{kA} \aleq K_{ij} \Big( \jj{+}{i}{k}(\jj{+}{j}{l}-\jj{-}{j}{l})+\delta^i{}_k \Pi^j{}_l ~,~
 \jj{+}{i}{k} \Pi^j{}_l  - \delta^i{}_k (\jj{+}{j}{l}-\jj{-}{j}{l}) \Big) k^l \\[2mm]
E_{AB} \aleq K_{ij} \left( \begin{array}{cc}
-(\jj{+}{i}{k} - \jj{-}{i}{k})(\jj{+}{j}{l}-\jj{-}{j}{l})  & -(\jj{+}{i}{k}-\jj{-}{i}{k}) \Pi^j{}_l \\[1mm]
\Pi^i{}_k (\jj{+}{j}{l}-\jj{-}{j}{l})  & \Pi^i{}_k \Pi^j{}_l
\end{array} \right) k^k k^l~.
\end{eqnarray}	

It is useful to investigate the low-energy properties of such $\sigma$-models when the kinetic terms for  $\dq{\phi,\chi},\dq{\prime}$ and $f$ flow to zero (\eg \cite{Hori:2000kt}). We now show two simple cases where the Lagrange-densities due to the same \kah potential (\ref{eq:kahlerpot}) in the (twisted)chiral scenario are compatible:
\begin{itemize}
 \item Quotient action: If there are no kinetic terms for $\Xi^{1,2}_\pm$ the two Lagrange-densities  are related by field redefinitions:
\begin{equation}
 \dq{\phi}_c = \dq{\phi}_t - iD_{[+}\hcc{-]} ~~,~~~\dq{\chi}_c = \dq{\chi}_t - iD_{(+}\hcc{-)} ~~,~~~\dq{\prime}_c = \dq{\prime}_t + iD_{[+}\hee{-]} ~~,~~~ A^c_\pm = A^t_\pm + 2 \hee{\pm}~.
\end{equation}
When integrating both $\Xi_\pm^A$ and $A_\pm$ we therefore obtain the same quotients.
\item T-duality: When adding a linear term that constrains the field-strengths to vanish
\begin{equation}
K \rightarrow K -\ft{1}{2} (\bbXh{L} V_L + \bbXBh{L} \bar{V}_L + \bbXh{R} V_R + \bbXBh{R} \bar{V}_R)~.
\end{equation}
we find \cite{Lindstrom:2007sq} that $\Xi_\pm^{1,2}$ are also constrained to vanish, which, in both cases, gives the action:
\begin{eqnarray}
\mathcal{L}\aleq K_{ij}(\jj{+}{i}{k} \jj{-}{j}{l}-\ft{1}{2} \Pi^i{}_k \delta^j{}_l
- \ft{1}{2} \delta^i{}_k \Pi_j{}^l ) \hcd{+}\hPh{k} \hcd{-}\hPh{l} \nn \\[1mm]
&+& i \dq{\phi}\big( K_i (\jj{+}{i}{j}+\jj{-}{i}{j}) k^j +i(\tilde{X}_L-\bar{\tilde{X}}_L+\tilde{X}_R-\bar{\tilde{X}}_R) \big) \nn\\[1mm]
&+& i\dq{\chi} \big(  K_i (\jj{+}{i}{j}-\jj{-}{i}{j}) k^j -i(\tilde{X}_L-\bar{\tilde{X}}_L-\tilde{X}_R+\bar{\tilde{X}}_R) \big) \nn\\[1mm]
&+& i\dq{\prime} \big(  K_i \Pi^i{}_j k^j 
-(\tilde{X}_L+\bar{\tilde{X}}_L+\tilde{X}_R+\bar{\tilde{X}}_R) \big) \nn\\[1mm]
&+& \ft{i}{2} f (\tilde{X}_L+\bar{\tilde{X}}_L-\tilde{X}_R-\bar{\tilde{X}}_R) \big)
\end{eqnarray}
\end{itemize}

\section{Conclusion}
In this paper we presented two possible candidates, which are unique to the abelian case, for \kah and superpotential terms for the large vector multiplet action where undesired higher derivative terms are removed by field redefinitions and choice of gauge invariants present in the \kah potential. We then write possible kinetic terms for the spinor invariants which include twisted conformal field theories and topological terms. This work concludes our presentation of the $\nZZ$ multiplets 
\cite{Lindstrom:2007vc,Lindstrom:2007sq,Lindstrom:2008hx} which is a step towards treating generalized \kah 
geometry on similar footing as torsion-free complex manifolds that arise naturally as target 
spaces for supersymmetric $\sigma$-models. Future work along this avenue may include a full treatment of the moment maps and generalized \kah quotients \cite{LRRvUZ}.

Another possible generalization resulting from these new gauge multiplets is the formulation of new gauge linear $\sigma$-models (GLSMs) with $H$-fluxes which can generalize results such as \cite{Adams:2006kb,Halmagyi:2007ft} or mirror symmetry (\eg \cite{Hori:2000kt,Hori:2003ic}). It is important, however, to note that the IR flow for the spinor invariants action of sec. \ref{ssec::kinetic_term} must be studied first as it may lead to new physical degrees of freedom in the effective theory.

\bigskip\bigskip
\noindent{\bf\Large Acknowledgments}:
\bigskip\bigskip
	
\noindent
It is a pleasure to thank Martin Ro\v{c}ek, Rikard von Unge and Maxim Zabzine for their help and 
encouragement, and the 2007 Simons workshop in Mathematics and Physics for providing a stimulating 
environment. This work was supported in part by NSF grant no.~PHY-0354776.

\end{document}